**Two liquid states of matter: A new dynamic line on a phase diagram**

V.V. Brazhkin[1], Yu.D. Fomin[1], A.G. Lyapin[1], V.N. Ryzhov[1] and Kostya Trachenko[2]

[1] *Institute for High Pressure Physics RAS, 142190 Troitsk Moscow region, Russia*

[2] *South East Physics Network and School of Physics, Queen Mary University of London, Mile End Road, London, E1 4NS, UK*

It is generally agreed that the supercritical region of a liquid consists of one single state (supercritical fluid). On the other hand, we show here that liquids in this region exist in two qualitatively different states: "rigid" and "non-rigid" liquid. Rigid to non-rigid transition corresponds to the condition $\tau \sim \tau_0$, where $\tau$ is liquid relaxation time and $\tau_0$ is the minimal period of transverse quasi-harmonic waves. This condition defines a new dynamic line on the phase diagram, and corresponds to the loss of shear stiffness of a liquid at all available frequencies, and consequently to the qualitative change of many important liquid properties. We analyze the dynamic line theoretically as well as in real and model liquids, and show that the transition corresponds to the disappearance of high-frequency sound, qualitative changes of diffusion and viscous flow, increase of particle thermal speed to half of the speed of sound and reduction of the constant volume specific heat to $2k_B$ per particle. In contrast to the Widom line that exists near the critical point only, the new dynamic line is universal: it separates two liquid states at arbitrarily high pressure and temperature, and exists in systems where liquid – gas transition and the critical point are absent overall.

# 1. Introduction

Our current understanding and discussion of basic states of matter such as solid, liquid and gas is illustrated by "temperature, pressure" ($T,P$) or "temperature, density" ($T,\rho$) phase diagrams. Crossing a line on such diagrams corresponds to thermodynamic phase transitions leading to qualitative changes of physical behavior of the system. Below we show that for one basic state of matter, the liquid phase, an equally important qualitative change of system behavior exists which is related to the change of its dynamics rather than thermodynamics. Consequently, we propose that all liquids have two qualitatively different states, and that a new additional line should be added to the phase diagram which separates the two states.

More specifically, a typical ($T,P$) diagram [Fig. 1(a)] implies that a liquid is separated from a gas by the boiling line ending at the critical point. The diagram further implies that only one single state (frequently called "supercritical fluid") exists for all pressures and temperatures above the critical point. On the other hand, we propose that an important qualitative change in a liquid behavior takes place on crossing our new line. Importantly, the new line extends for arbitrary values of pressure and temperature above the critical point [Fig. 1(a)]. In addition, the new line is not related to the critical point from the physical perspective, and therefore exists in systems where the liquid-gas transition is absent overall, as is the case in some soft matter subjects with short-range attractive forces as well as in model soft-sphere system [Fig. 1(b)].

We begin our discussion with the work of J. Frenkel [1], who provided a microscopic description of Maxwell phenomenological viscoelastic theory of liquid flow [2], by introducing liquid relaxation time $\tau$: $\tau$ is the average time between two consecutive atomic jumps in a liquid at one point in space. Each jump can approximately be viewed as a jump of an atom from its neighboring cage into a new equilibrium position, with subsequent cage relaxation. These atomic jumps give a liquid its ability to flow. The relaxation time $\tau$ is a fundamental flow property of a liquid, and it defines liquid viscosity $\eta$ and diffusion coefficient $D$.

The above picture implies that the motion of an atom in a liquid consists of two types: quasi-harmonic vibrational motion around an equilibrium position as in a solid and diffusive motion between two neighboring positions, where typical diffusion distances exceed vibrational distances by about a factor of ten [Fig. 2(a,b,c)]. Therefore, atomic motion in a liquid combines both elements of the short-amplitude vibrational motion as in a solid and the large-amplitude ballistic-collisional motion as in a gas. Our main proposal is that the point at which the *solid-like motion ceases*, leaving only the gas-like motion, marks the change of most important properties of a liquid.

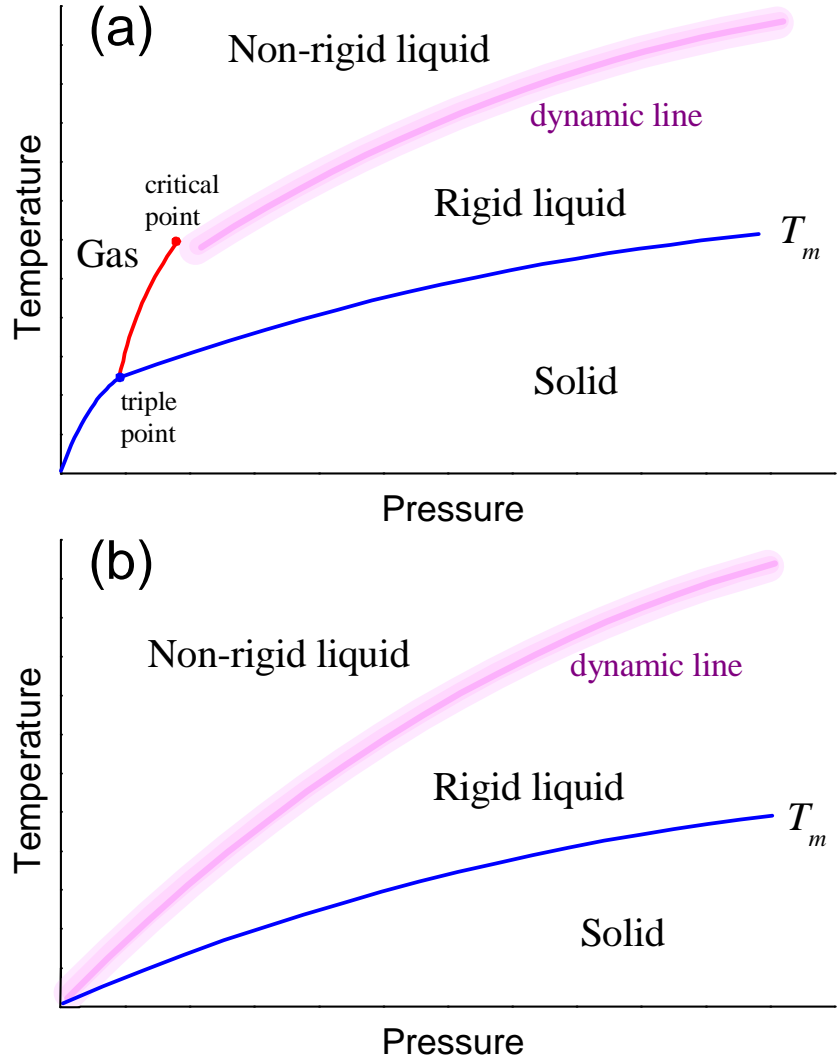

Fig. 1. Pressure-temperature phase diagrams of an ordinary substance (a) and a system without the boiling line and liquid-gas critical point (b). In both cases, there exists a dynamic line separating rigid and non-rigid liquids.

The value of $\tau$ decreases with temperature increase, spanning many orders of magnitude. On the other hand, the minimal (Debye) vibration period, $\tau_0$ ($\tau_0 \approx 0.1$–$1$ ps), is weakly temperature-dependent, and is mostly defined by interactions in a given system. At certain high temperature the solid-like vibration character ceases [Fig. 2(a,d,e)]. This point is reached when $\tau$ becomes comparable to $\tau_0$:

$$\tau \sim \tau_0. \qquad (1)$$

In the following discussion, we consider $\tau$ as the average time it takes an atom to move the average inter-particle distance $a$. Then, $\tau$ quantifies the motion envisaged by Frenkel, where an atom jumps distance $a$ during time $\tau$ between two equilibrium positions at low temperature as well as the motion at high temperature where two equilibrium positions are absent altogether and the motion between collisions is ballistic as in a gas. Here, $\tau$ is the time between collisions.

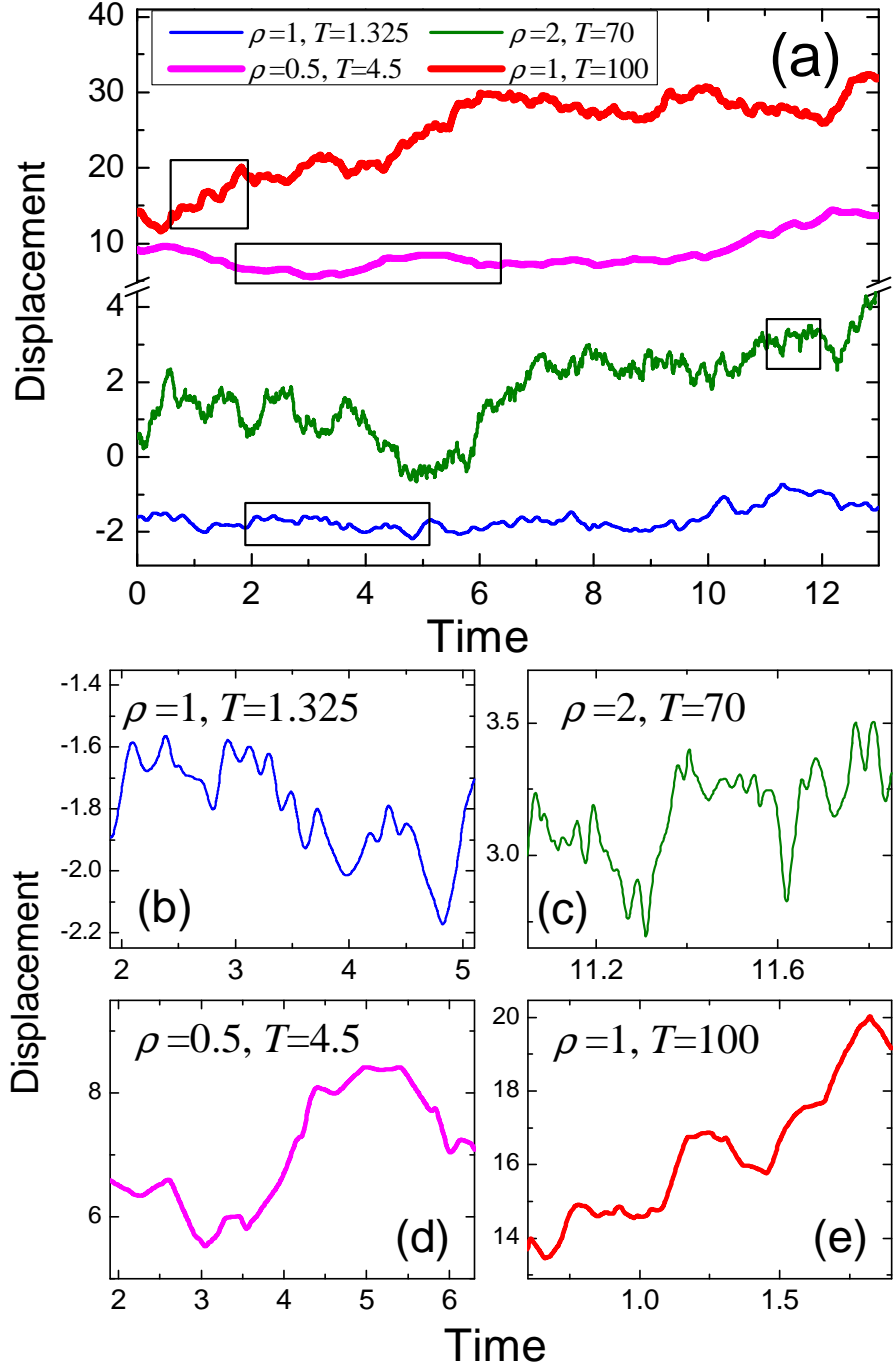

Fig. 2. Examples of particle trajectories ($x$ coordinate) for the Lennard-Jones (LJ) liquid at different conditions, presented in LJ units ($\rho_c=0.314$, $T_c=1.31$). Panels (b)-(e) show selected fragments from (a). Cases (b) and (c) correspond to the rigid state where vibrations are present; (d) and (e) correspond to collisional motion in the non-rigid state.

We note that for $\tau>\tau_0$ where dynamics is mostly vibrational, the atomic jumps take place by activation over the barrier created by the potential energy of interaction [1]. Therefore, the transition from solid-like vibrations to continuous gas-like ballistic motion takes place when kinetic energy *K* of a particle becomes comparable to potential energy of their interaction. Hence, condition (1) implies crudely

$$3k_BT/2 \sim E_{\text{pot}}. \tag{2}$$

For most substances, the kinetic energy to the potential energy $K/E_{pot}$ ratio at melting temperature is significantly lower than 1. In this case the particles are primarily within the region of the action of the potential and, at melting, relatively long-lived regions with a well-defined vibration spectrum and short-range order remain. We will elaborate on this below.

Condition (1) is achieved on a certain new line on the phase diagram. Below we show that it leads to important qualitative changes of the system behavior, as witnessed by the change of its elastic, structural, dynamic, diffusive and thermodynamic properties when the system crosses the new line.

## 2. Solid-like elastic and structural properties

An important realization from the introduction of relaxation time is that if observation time is smaller than $\tau$, the local structure of a liquid does not change, and is the same as that of a solid. This enabled Frenkel to predict that a liquid should maintain solid-like shear waves at all frequencies $\omega > 2\pi/\tau$ [1]. This prediction was later confirmed experimentally for different kinds of liquids [3-5].

The maximum oscillation frequency available in the system is $\omega_0 = 2\pi/\tau_0$. Therefore, solid-like shear waves exist in the range $2\pi/\tau < \omega < 2\pi/\tau_0$. Consequently, condition (1) ($\tau \sim \tau_0$) corresponds to the complete loss of shear waves and therefore to the loss of shear resistance, or rigidity, at all frequencies existing in the system.

We note that condition (1) and other conditions considered below correspond to approximate equality. Moreover there are wide distributions of the $\tau$ and $\tau_0$ values, and we consider their average values. Nevertheless in all liquids there are definite *T,P*–conditions corresponding to the loss of transverse-like vibrations in the excitation spectra, hence we can speak about definite line instead of wide crossover.

The ability of liquids to flow is often associated with zero rigidity, or shear resistance, that markedly distinguishes liquids from solids. However, this implies zero rigidity at small frequencies only, whereas at larger frequency, a liquid supports shear stress. On the other hand, condition (1) marks the qualitative change, from the physical point of view, of system elastic properties: shear resistance is lost completely, at all frequencies available in the liquid. Therefore, condition (1) marks the crossover between a *"rigid" liquid*, where rigidity exists in a certain frequency range and a *"non-rigid" liquid* which can not sustain rigidity at any frequency.

We note that crossover (1) has important implications for the ability of liquids to undergo phase transitions, an emerging and fast-developing area of research [6-8]. When $\tau>\tau_0$ at low temperature, local structure, or definite short-range order structure (SROS), of the liquid remains unchanged. In this case, pressure, temperature changes can induce a phase transition in a liquid, accompanied by the change of the SROS and dynamics [6-8]. On the other hand, when $\tau<\tau_0$ at high temperature, the only random packing structure type can exist because atoms are in the continuous gas-like state of motion. In this case, pressure can only induce a change of density but not a phase transition with change of SROS. Therefore, the proposed dynamic line demarcates the regions on the phase diagram where phase transitions in a liquid operate.

**3. Dynamics and sound propagation**

The qualitative change in atomic dynamics, defined by Eq. (1), has important consequences for atomic dynamics, sound propagation and diffusion. Lets approach the liquid from low temperature where $\tau>\tau_0$. The speed of sound in a liquid or a solid, $V_s$, is defined from the dispersion relation, $\omega=V_s k$. Using linear Debye approximation and taking maximum frequency $\omega$ as Debye frequency $\omega_0=2\pi/\tau_0$, and $k_{max}=\pi/a$ gives $V_s=2a/\tau_0$. Lets now approach the liquid from high temperature where oscillatory motion is lost, and recall that $\tau$ is the time between two consecutive collisions over distance $a$. Then, $V_{th} \sim a/\tau$, where $V_{th}$ is particle thermal velocity. Therefore, condition (1) implies

$$V_s \sim 2V_{th}. \tag{3}$$

In condensed phases such as solids and liquids, the speed of sound is primarily determined by the interactions between atoms: the sound velocities are given by elastic moduli. In dense liquids, moduli vary insignificantly on the isochors [9], and sound velocities are weakly temperature-dependent. On the other hand, the thermal velocity of a classical particle increases with temperature without bound. Therefore, a temperature range must exist where the speed of sound and thermal velocity become comparable in magnitude. The physical meaning of condition (3) is that particles cease to feel elastic resistance of the medium, and start moving in a ballistic way.

We note that Eq. (3) is based on the same physical grounds as in Eq. (2). Indeed, the speed of sound is governed by the elastic moduli, which are in turn proportional to the potential energy of the system per unit volume. However the proportionality coefficient in (2) may be significantly (several times) different from 1.

Another interesting consequence of Eq. (1) is related to the phenomenon of "high frequency sound" or positive dispersion of sound velocity, which is the increase of the speed of sound at high frequencies. Frenkel predicted [1] that this effect should exist for frequencies

$\omega > 2\pi/\tau$. If, as he argued, shear waves kick in at frequency $2\pi/\tau$, the speed of sound increases from $(B/\rho)^{1/2}$ to $[(B+4G/3)/\rho]^{1/2}$, where $B$ is a bulk modulus and $G$ is a shear modulus of a liquid, because $G$ becomes non-zero at this frequency. This viscoelastic model was later developed in detail, including memory function formalism, nonlocal mode coupling theory etc [10-14]. Following the prediction, the "high frequency sound" was observed in numerous experiments, receiving particular attention since the development of inelastic X-ray techniques [15-18]. We now observe that the proposed crossover (1) marks the point at which the positive dispersion disappears completely, because, as discussed above, this crossover corresponds to the complete loss of shear waves that can exist in a liquid.

### 4. Diffusion and viscosity

The change of the character of atomic diffusion in the liquid at the crossover (1) occurs at a particular value of diffusion constant $D^*$. $D$ can be estimated as $D=a^2/6\tau$. When $\tau \sim \tau_0$ at the crossover (1), we have

$$D = D^* \sim a^2/6\tau_0. \qquad (4)$$

Taking $a \sim 1$ Å and $\tau_0 \sim 0.1$ ps gives $D^* \sim 10^{-8}$ m$^2$/s. The condition $D=D^*$ provides good estimation of the dynamic line defined by condition (1) because it is fairly insensitive to the increases of pressure and temperature. Indeed, both $a^2$ and $\tau_0$ decrease with pressure only slightly, and their ratio becomes even less sensitive to pressure and temperature. We note that $D^*$ value is consistent with the experimental values of diffusion near critical point, $D_c$ [9]. Indeed, in the vicinity of the critical point, liquids are known to loose their elastic properties and relaxation process changes its nature from activation to collisional one [19,20], hence the near equality of $D^*$ and $D_c$ is not surprising in our picture. Thus, for crude estimation of the crossover (1) at moderate pressures, we can use the condition

$$D \sim D_c. \qquad (5)$$

In this sense, the proposed dynamic line starts not far from the end of liquid-gas transition. However, it is important to stress that all basic conditions above, (1), (2), (3) and (4) are not related to liquid-gas transition and to the existence of critical point from the physical point of view, and continue to operate in systems where the critical point is absent altogether. We will further comment on this below.

Importantly, condition (1) corresponds to the crossover between two different qualitative temperature dependencies of diffusion $D$ and viscosity $\eta$. Indeed, when $\tau > \tau_0$ at low temperature, $\tau \sim \exp(U/T)$, where $U$ is the activation barrier. Then, $D \sim a^2/\tau \sim \exp(-U/T)$. On the other hand, when $\tau < \tau_0$ at high temperature, $\tau$ quantifies thermal motion as discussed above: $\tau \sim 1/V_{th} \sim 1/T^{1/2}$, giving

$D \sim T^{1/2}$ for a low density gas or $D \sim T^\alpha$, where α is almost constant for a dense fluid [21]. Therefore, condition (1) gives the crossover of $D$ from exponential to power-law temperature dependence. Similarly to $D$, temperature behavior of viscosity $\eta$ changes at the crossover (1). Indeed, when $\tau > \tau_0$, $\eta$ almost exponentially decreases with temperature, which can be seen most easily by applying the Maxwell relation $\eta = G_\infty \tau$, where $G_\infty$ is the instantaneous shear modulus having weak temperature dependence in comparison with the exponential decrease of $\tau$. On the other hand, when $\tau < \tau_0$, $\eta \sim T^{1/2}$. This follows from applying either the Stokes-Einstein-Debye relationship, $\eta \sim T/D$ or the Maxwell relationship $\eta = G_\infty \tau$ where $\tau \sim 1/T^{1/2}$ from above and recalling that $G_\infty$ is proportional to kinetic $\sim T$ term in this regime [22]. We therefore conclude that condition (1) corresponds to the qualitative change in the temperature behavior of viscosity, as it crosses over from the exponential decrease at $\tau > \tau_0$ to the power increase at $\tau < \tau_0$.

## 5. Thermal energy and specific heat

Experimentally, constant-volume specific heat of liquids at ambient pressure decreases from about $3k_B$ per particle around the melting temperature to about $2k_B$ at high temperatures [9,23]. Further decrease of specific heat with temperature increase up to gas-like values $3k_B/2$ is observed at high pressures in supercritical region [9]. This behavior was quantitatively on the basis of decreasing contribution of shear modes to liquid energy with temperature [24]. In this model, the liquid thermal energy per atom is

$$\frac{E}{N} = k_B T \left[ 3 - \left(\frac{\tau_0}{\tau}\right)^3 \right]. \tag{6}$$

According to Eq. (6), when $\tau$ considerably exceeds $\tau_0$ at low temperature, liquid energy is close to $3k_B T$ per atom, giving the Dulong-Petit value of specific heat of $3k_B$. When $\tau$ approaches $\tau_0$ at high temperature, liquid energy becomes $2k_B T$ per atom, giving the specific heat of $2k_B$, consistent with the experimental results. At this temperature, shear waves are completely lost at all frequencies, and longitudinal modes only contribute to the heat capacity. Therefore, the crossover from a rigid to a non-rigid liquid at $\tau \sim \tau_0$ is accompanied by the decrease of the specific heat from its solid-state value to the value of $2k_B$:

$$c_V \sim 2k_B. \tag{7}$$

Eq. (7) gives the minimal value of specific heat that a rigid liquid can possibly have. Further decrease of heat capacity corresponds to the loss of longitudinal modes as the temperature is increased until the gas-like state is reached with $c_V = 3k_B/2$.

## 6. Methods

We have studied Lennard-Jones (LJ) liquid and two Soft Spheres (SSp) liquids with $n=12$ and $n=6$ in a very wide range of parameters, from temperature $T=0.6$ (well below the critical point) to $T=100.0$ and densities from 0.1 to 2.7 for LJ liquid, $T=0.1–100$ and density in the range 0.1–3.33 for SSp liquid with $n=12$ and $T=1.0–10$ and density in the range of 0.5–4.5 for SSp liquid with $n=6$. We also have studied SSp liquids with high values of $n$ up to 100 for the calculation of $E_{pot}$ and $c_V$. An essential property of soft spheres is that the phase diagram corresponds to the equation $\gamma = \rho\sigma^3 \left(\frac{\varepsilon}{k_B T}\right)^{3/n} = \text{const}$. The parameters $\gamma$ for $n=12$ and $n=6$ systems were taken $\gamma_{12} = 1.15$ and $\gamma_6 = 1.22$ [25] (see also **Supporting Information**).

To calculate the Debye period $\tau_0$, we analyzed many particle trajectories. At low temperature, $\tau_0$ was obtained as the average time of oscillatory motion at one equilibrium position of an atom before jumping to another position. At high temperature, the oscillations disappear and the motion of particle consists of collisional movement only. The value of $\tau$ was calculated as time it takes a particle to move the distance close to the average inter-particle separation. Then, it can be estimated as $\tau = \frac{\rho^{-2/3}}{6D}$, where $D$ is the diffusion coefficient. $D$ was calculated from the long-time limit of mean square displacement using the Einstein relation (see also **Supporting Information**). The values of $\tau_0$ and $\tau$ as well as the temperature of disappearance of solid-like vibration motion under heating can be also extracted from the analysis of self-intermediate scattering function $F_s(q,t)$ [11] (Fig.3) and mean square displacement $<x^2>$ at different temperatures.

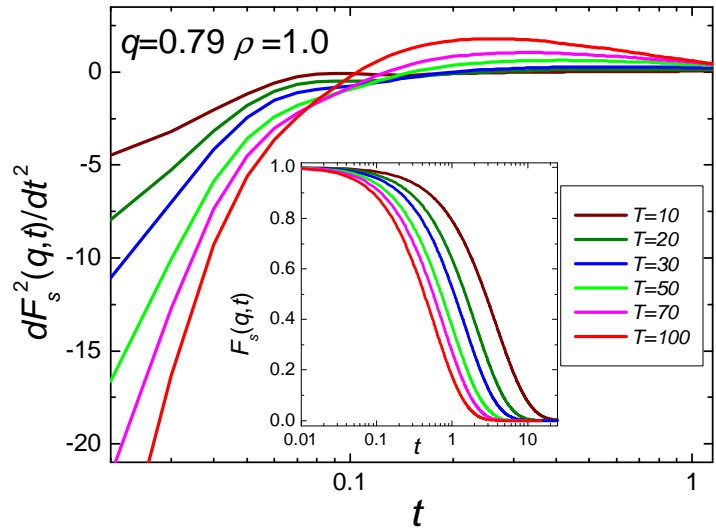

Fig. 3. Autocorrelation scattering functions of the LJ liquid (in the inset) and their second-order time derivatives. Temperature is pointed in LJ units.

All other quantities discussed in the paper were calculated in standard way (see **Supporting Information**). We also use the data of potential energy and equation of state for LJ fluid from [26], and the melting line of LJ system from [27].

Thermodynamic and dynamic data for real fluids (Ar, Ne, and $N_2$) were taken from [9].

## 7. Evidence from molecular dynamics simulations and experimental data

Having discussed the main physical properties that change at the crossover (1), we now provide numerical and experimental evidence supporting our proposal. We have drawn the lines determined by Conditions (1), (2), (3), (5), and (7), for real substances (Ar, Ne and $N_2$) and model particle systems with Lennard-Jones (LJ) and soft-sphere potentials (SSp) (Figs. 4 and 5). It is well known that LJ potential adequately represents the behavior of molecular and rare gas liquids whereas SSp systems describe a behavior of real substances at very high pressures where an attractive term of the inter-particle potential can be neglected. All these systems are strongly correlating according to recent classification [28]. We have calculated points on the phase diagram that correspond to strict equalities $\tau=\tau_0$, $3kT/2=E_{pot}$, $V_s=2V_{th}$, $D=D_c$, and $c_V=2k_B$, hence moderate differences between the lines positions are not unexpected.

According to Fig. 4, the position of the crossover line (Condition 1) for Lennard-Jones system agrees with the line determined by Condition (5) at moderate pressures $P<10 - 10^2 P_c$ and with the lines determined by Conditions (3) and (7) at high pressures $P>10P_c$. At low pressures, the lines determined by Conditions (3) and (7) shift from the crossover line (1) due to critical point anomalies and loss of Debye approximation at low densities. The Condition (2) is not directly based on the Condition (1) and proportionality coefficient in (2) differs from 1 significantly (for LJ particles $3kT/2 \approx 5E_{pot}$, for SSp particles ($n=6$) $3kT/2 \approx 0.3E_{pot}$). For a soft-sphere system, the lines determined by the Conditions (1), (3), and (7) match well over the entire pressure range (Fig. 5), as no critical point and associated anomalies exist for this system.

In addition to model systems, we find good agreement between the theoretical predictions and experimental data for liquid Ar, Ne, and $N_2$ (Fig. 4). In particular, we observe a good match between the region of the disappearance of the positive dispersion of sound velocity in liquid Ar [16] and $N_2$ [18] and the dynamic line (Fig. 4). The qualitative change of the temperature dependencies of viscosity $\eta$ also occurs near crossover line (Fig. 6).

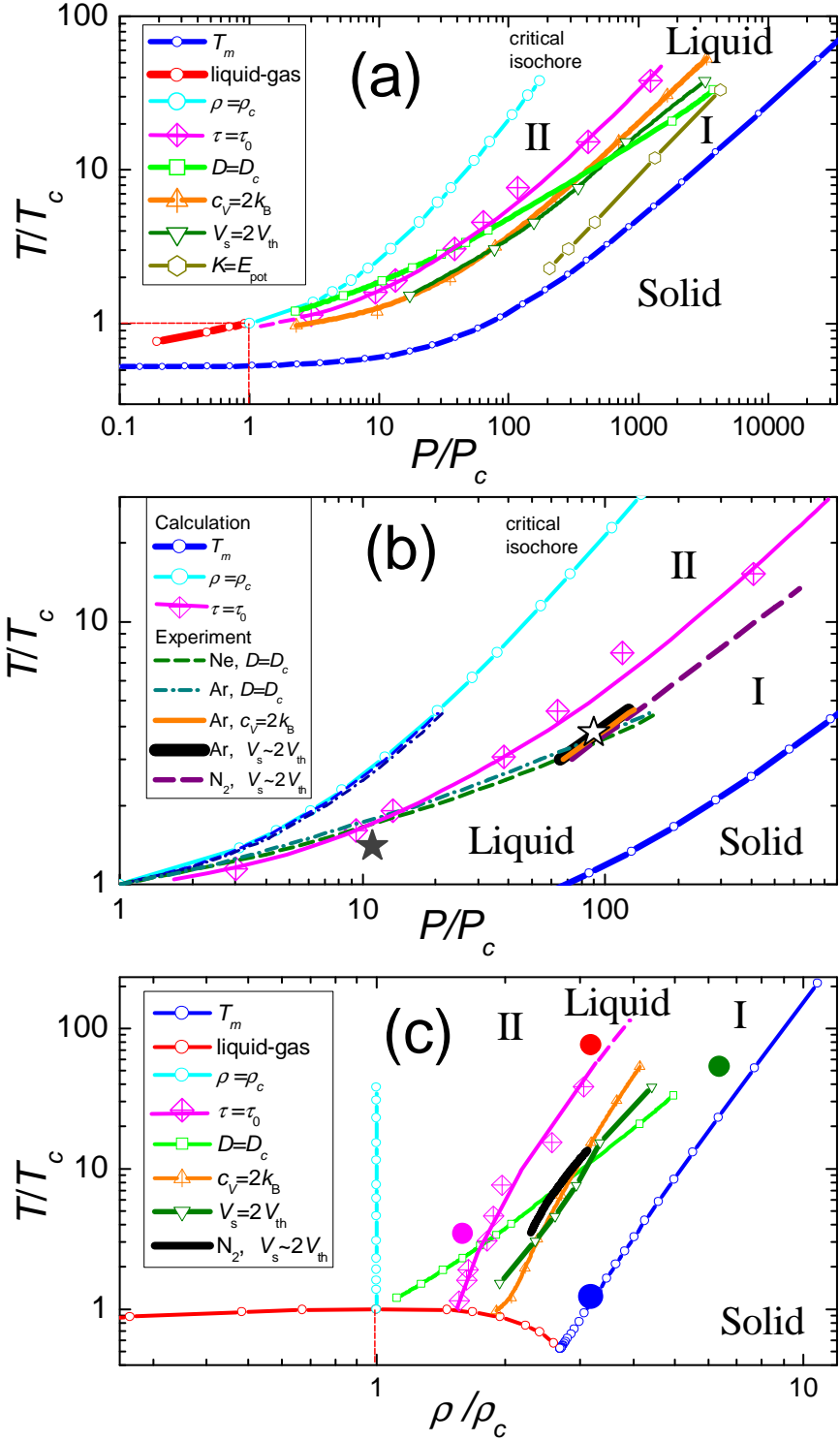

Fig. 4. (*T*,*P*) phase diagram [(a) and (b)] and (*T*,*ρ*) phase diagram (c) of the LJ liquid in the relative critical coordinates. Panels (a) and (c) present calculated lines defined by different criteria (see the text). Panel (b) presents some experimental data from [9]. Stars in the panel (b) correspond to known experimental points where liquid looses shear waves and positive dispersion (open symbol for Ar [16] and solid symbol for $N_2$ [18]). Experimental critical isochors are also shown in panel (b) (dashed for Ne and dashed-dotted for Ar). Experimental data for criterion (3) for nitrogen and points (solid circles) from the simulation of the LJ system in Fig. 2 (with the same color) are shown in panel (c). In all cases number I correspond to rigid liquid and II – to non-rigid one.

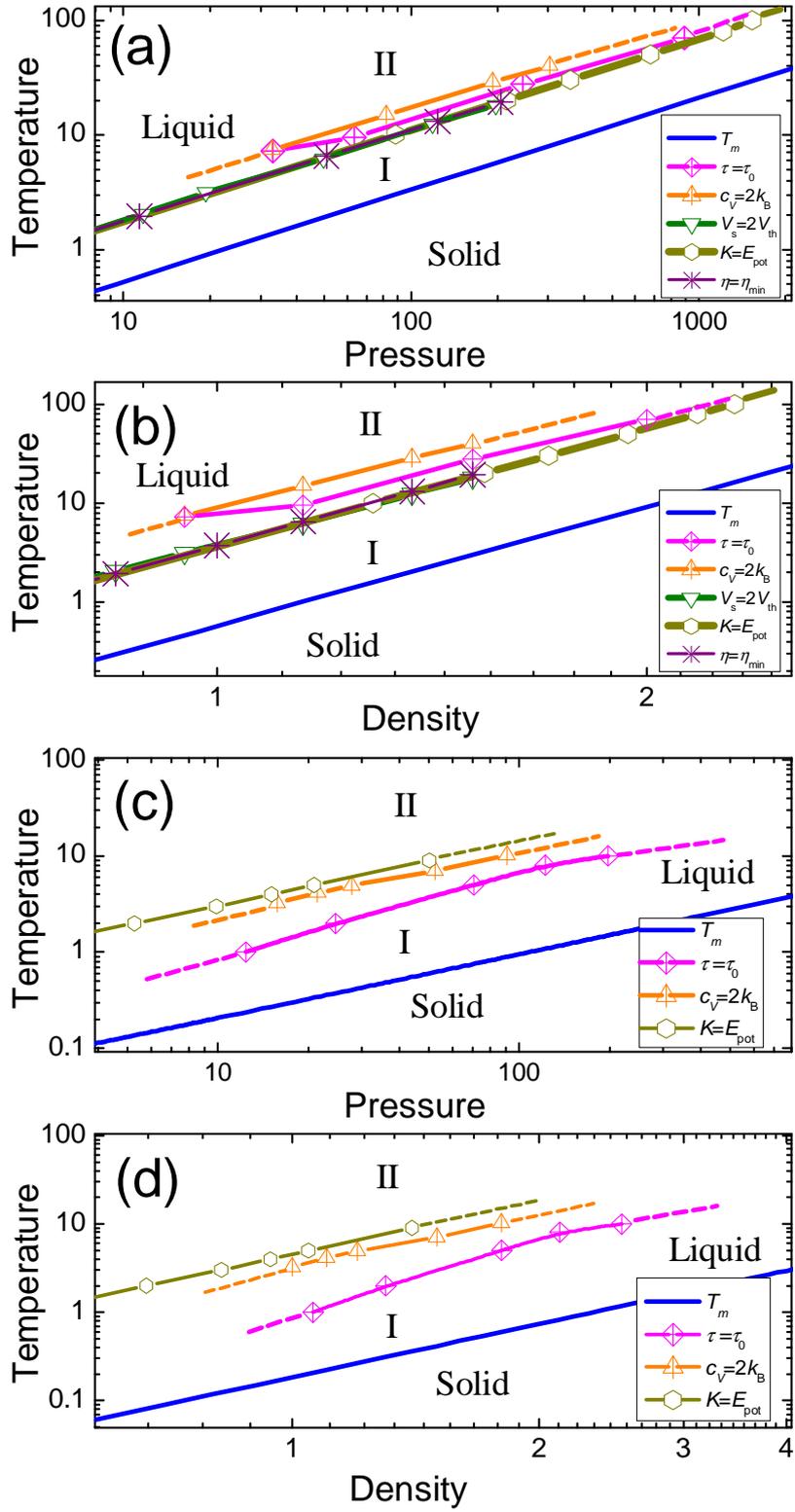

Fig. 5. ($T,P$) and ($T,\rho$) phase diagrams of the simulated soft-sphere systems with $n=12$ [(a) and (b)] and $n=6$ [(c) and (d)]. This figure presents calculated lines defined by different criteria including that for the minimum of viscosity $\eta$ on panel (b).

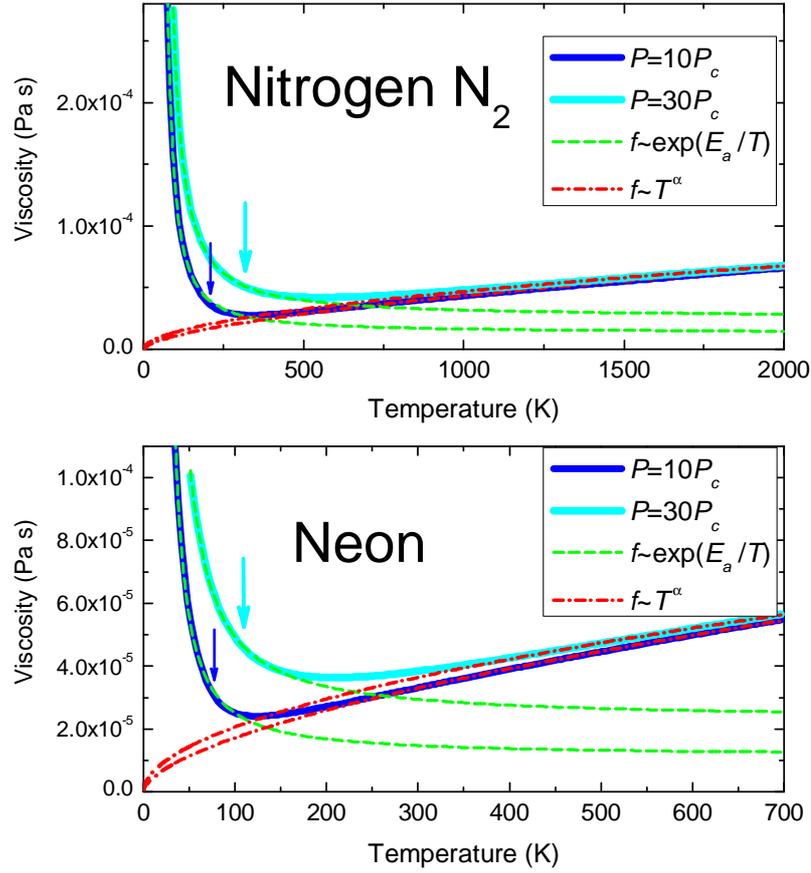

Fig. 6. Experimental [9] isobaric temperature dependences of viscosity of nitrogen and neon with asymptotes at low (exponent of inverse temperature) and high (power law) temperatures. Arrows indicated the corresponding temperatures of condition (1), $\tau=\tau_0$. $\alpha=0.59$ at $P=10P_c$ and 0.52 at $P=30P_c$ for Ne and similarly 0.59 and 0.53 for $N_2$.

## 8. Dynamic line and melting line

Interestingly, all crossover lines correspond to density increase with increasing temperature: the relation $\rho \sim T^k$ is met, where $k \sim 0.25$ for liquid $N_2$, Lennard-Jones and soft-sphere system with $n=12$ (Figs. 4 and 5). Note that for any system of particles with uniform potential, there are scaling relations for physical values [29]. In particular, $\rho^{n/3}/T=$const along the melting line for the soft-sphere system, giving $\rho \sim T_m^{1/4}$ for the melting temperature $T_m$ for $n=12$ [29]. The similarity of the dependence of $\rho(T)$ for the melting and dynamic lines implies similar scaling relations for the dynamic line. This point will be discussed in details elsewhere. As a result, the region of the "rigid" liquid does not, under any pressures, disappear (Figs. 4 and 5). Consequently, the dynamic line continues for arbitrarily high pressures and temperatures.

It should be mentioned that the "softer" the repulsion potential, the wider the region of existence of a rigid liquid (Fig. 5). We have calculated the ratio between the kinetic energy of the particles and the potential energy of the particle interaction $K/E_{pot}$ along the melting curve for soft sphere systems with different repulsion coefficient $n$ (see Fig. 7). This ratio varies from 0 for $n=3$ to infinity for $n \to \infty$ (hard spheres). The value of $K/E_{pot}$ near the melting curve is considerably larger than 1 for only very large coefficients $n>30$. For even larger values of $n$, the dynamic line can "hide behind" the melting curve. Particularly, the condition (7) $c_V \sim 2k_B$ is fulfilled near the melting curve with $n\sim 60$. Thus, for $n > 50$-$60$, the dynamic line falls within the region of crystal stability and separates a low-temperature, almost harmonic state of the crystal and a high-temperature, strongly anharmonic state. In the high-temperature state, a particle moves most of the time in a ballistic way outside the region of the action of the potential similarly to the behavior of the particles of a hard sphere crystal, for which the potential energy is equal to zero, and $c_V = 3/2 k_B$. In this case the condition (1) $\tau \sim \tau_0$ practically implies that a particle in the crystal moves comparable time almost harmonically and ballistically.

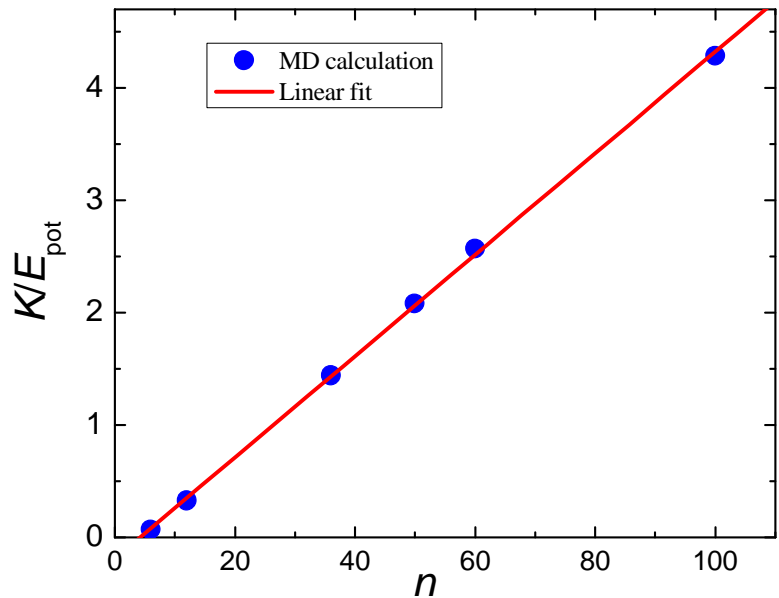

Fig. 7. Kinetic-to-potential energy ratio for the SSp liquid at the melting curve (it is pressure independent) as function of the soft-sphere potential exponent $n$.

Formally, the particle motion for the hard sphere system in the overcooled region can also tentatively be split into 2 types: a small amplitude motion inside the cages and rare jumps over large distances. Despite similarity of this picture to the behavior of particle trajectories in normal liquid, there are considerable distinctions. The hard spheres motion inside the cages is of a purely collisional, absolutely nonharmonic character, and the jumps only involve geometry restrictions rather than overcoming the activation barrier. In this case $c_V \sim 3/2 k_B$, the speed of sound and thermal speed are not independent quantities, the structure corresponds to the random packing of spheres, and all transverse-like excitations damp at wavelengths of the order of one wavelength. Therefore, it is appropriate to treat the hard sphere liquid as a non-rigid liquid at all temperatures.

The above results are important for understanding the difference between liquid-glass transition and a jamming transition, the question that has been widely discussed and debated. For very large values $n > 100\text{-}150$, the glass transition in an undercooled metastable melt occurs in a non-rigid liquid state and, like the hard-sphere glass transition, is governed by geometrical confinement conditions, and there is no temperature region with activation behavior with almost constant activation energy.

It is known that for particle systems with a narrow region of the action of the potential, the liquid-gas equilibrium line in the equilibrium phase diagram is absent; instead, there is a solid state crystal-crystal isostructural transition ending in a critical point [30-32]. In this case the discussed dynamic line, separating the harmonic and anharmonic states of the crystal, is in fact a continuation of the above isostructural transition into the supercritical region. Thus, for some colloidal and macromolecular systems, for which the boiling line is absent, the dynamic line can also lie in the region of stability of a solid phase.

### 9. Dynamic line and Widom line

It is interesting to note the recent attempts [16, 17] to link the change in the excitation spectrum to the "thermodynamic" continuation of the boiling curve, the so-called Widom line, the line of the maxima of thermodynamic properties in the vicinity of critical point [33]. In [16] the Widom line was discussed for heat capacity only and, notably, experimental data was extrapolated to very higher pressures to give an extrapolated thermodynamic line. From a physical point of view, this extrapolated line is qualitatively different from the dynamic line proposed here. Indeed, the proposed dynamic line is not related to the extrapolation of the boiling curve, and exists in systems where liquid-gas transition and the correspondent Widom line are absent altogether, including in model soft-sphere system, some colloidal systems, macromolecules and so on [30-32]. In addition, there are several other important differences. First, we the maxima of heat capacity $c_P$ become smeared at $T/T_c > 2.5$ and $P/P_c > 15$ [9] and the extrapolation into the high pressure range ($P/P_c \sim 100$) made in [16] is not physically meaningful. Second, we have calculated compressibility $\beta_T$, expansion coefficient $\alpha_P$, heat capacity $c_P$ and density fluctuations $\varsigma$ for the Lennard-Jones particle system. The results, together with experimental data for Ar and Ne, are summed up in Fig. 8. We observe that the thermodynamic continuation of the boiling curve gives a single line within 10 % departure in temperature from the critical point only. Upon further departure, it represents a rapidly widening bunch of lines instead of one single line, implying further that the extrapolation of the thermodynamic Widom line is not unique. The maximum of $\beta$ becomes smeared at $T > 1.1 T_c$ and its line is not shown in Fig. 8. Third, we observe that the lines of maxima of all calculated properties rapidly decrease in magnitude and become smeared at $T \sim (2-2.5) T_c$ and $P \sim (10-15) P_c$, and therefore can not be meaningfully extrapolated to higher pressures and temperatures. This is in contrast to the proposed dynamic line which, as discussed above, exists for arbitrarily high pressures and temperatures. Finally, we note that apart from the line of the maxima of the heat capacity lying close to the critical isochore, the lines of thermodynamic anomalies correspond to a decrease in the density with temperature increase, in strong contrast to the dynamic line (Fig. 8).

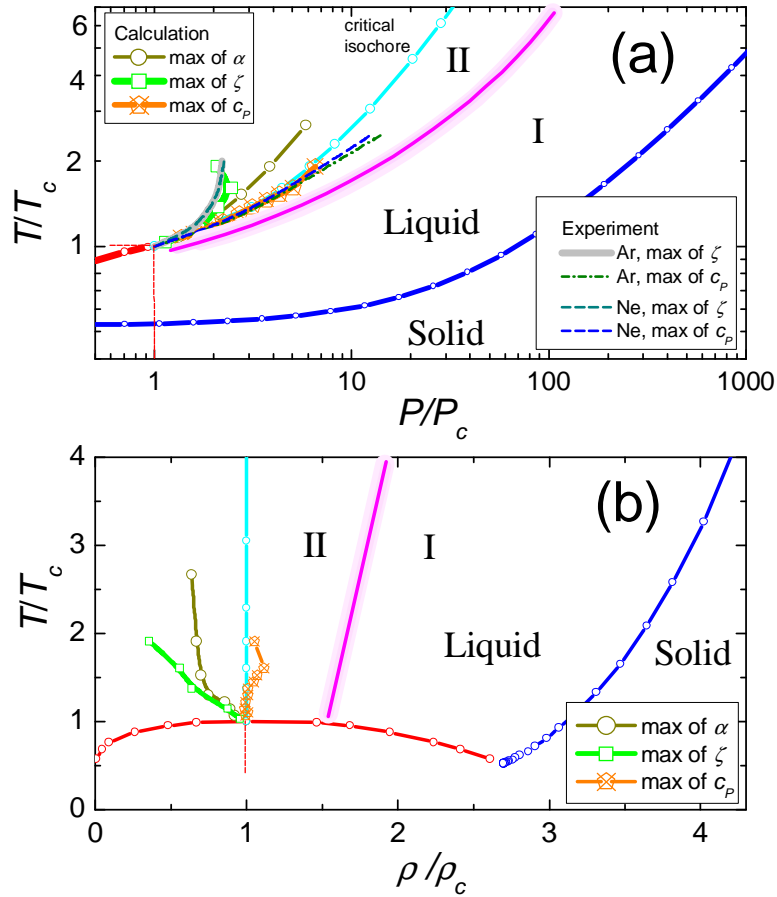

Fig. 6. ($T,P$) and ($T,\rho$) phase diagrams of the simulated LJ liquid (the basic notations are the same as in Fig. 3) with the dynamic line $\tau=\tau_0$ and calculated detectable points of maximums for thermal expansion $\alpha$, fluctuations $\zeta$, and isobaric thermal capacity $c_P$. The similar experimental data for Ar and Ne [9] are shown in panel (a).

## 10. Conclusions

One can mention that in most of real metallic, covalent and ionic liquids the dynamical line lies at extreme experimental conditions, e.g. ~10 GPa and ~$10^4$ K, and only shock-wave experiments can be used for its study. However for molecular and rare-gas liquids the line is situated at "static pressures" experimental condition, e.g. for Ne the condition (1) at $P$~3 GPa should take place at $T$~1000K (approximately 5 times higher than the melting temperature). For many soft matter subjects like colloidal systems, macromolecules etc the dynamical line also lies in the accessible range of pressures and temperatures (sometimes, is in the stability field of solid phases).

We propose to call the dynamic line defined by Eq. (1) "Frenkel line", to honor the contribution of J. Frenkel to the area of liquid dynamics. The contribution started from the microscopic definition and discussion of liquid relaxation time $\tau$. On the basis of this property, Frenkel made a number of important predictions regarding flow, relaxation as well as elastic and phonon properties of liquids that subsequently formed the microscopic basis of what is now known as "visco-elastic" picture of liquids [1]. One should mention that Frenkel' ideas were significantly refined and successfully used for last 20 years by D.C. Wallace group to calculate the thermodynamic and dynamic properties of liquid [34,35].

The proposed Frenkel line separates a "rigid" liquid where solid-like shear waves exist and diffusion regime is jump-like and activated as in a solid, from a "non-rigid" liquid where no shear modes exist and diffusion is collisional as in a gas. This line can be mapped in future experiments using several conditions for liquid properties that we discussed, including the disappearance of SROS peculiarities, the disappearance of positive dispersion of sound velocity, $D=D_c$, $V_s/V_{th}=2$, and $c_v(T)=2k_B$.

Acknowledgments:

The authors wish to thank S.M. Stishov, H.E. Stanley and P.F. McMillan for valuable discussions. The work has been supported by the RFBR (11-02-00303, 11-02-00341 and 10-02-01407) and by the Programs of the Presidium of RAS. K.T. is grateful to EPSRC.

**Supporting Information**

System size in the simulations varied depending on the density reaching 4000 particles at the highest densities. The cut-off radius was set to $2.5\sigma$ for LJ and SSp with $n=12$ and half of the box size for SSp with $n=6$. The equations of state were integrated by velocity Verlet algorithm. The temperature was kept constant during the equilibration by velocities rescaling. When the equilibrium was reached the system was simulated in NVE ensemble. The usual equilibration period was 1.5 million steps and the production run – 0.5 million steps where the time step $dt=0.001$ LJ units.

The soft spheres system was simulated in NVE ensemble. The system consisted of 1000 particles, the time step was 0.0005. The equilibration and production periods were 3.5 million and 0.5 million step, respectively. The simulations and computation of properties were made in the same way as for LJ system.

We note that $\tau_0$ can be obtained in the low temperature limit where the oscillations are well-pronounced while $\tau$ can be calculated from the diffusion value both at low temperatures and at high temperature limit where there is a ballistic-collisional regime. It makes necessary to compute the values of $\tau_0$ and $\tau$ along the same isochors at different temperatures and extrapolate the data to see the cross points. The accuracy of the estimation of the temperature of crossover $\tau=\tau_0$ is about ±20–30 %; the errors in the calculations of other values is less then 10 %.

The kinetic energy is obtained as $K/N=3k_BT/2$, and potential energy as the total energy of interaction. In the case of LJ liquid there is the complication associated with attractive part and sign change of potential energy at compression. That's why we estimated the potential energy as $E_{pot}=E_{LJ}(V)-E_{LJ}(V_0)+P_0(V_0-V)$ where $V_0$ corresponds to the volume of minimal possible (negative) pressure $P_0$ at zero temperature. Infinite-frequency shear modulus $G_\infty$ was calculated as in [22], where it was shown that $G_\infty$ for pair-potential systems can be obtained from radial distribution function. The bulk modulus $B = \beta^{-1} = \rho\left(\frac{\partial \rho}{\partial P}\right)_T^{-1}$ was obtained from equation of state. The longitudinal and transverse sound velocities were calculated as $V_{sl}=(B/\rho)^{1/2}$ and $V_{st}=(G_\infty/\rho)^{1/2}$. The heat capacities at constant volume were obtained by differentiating the internal energy at isochors $c_V = \left(\frac{\partial U}{\partial T}\right)_V$. Having the dependence of isochoric heat capacity on temperature along an isochors we found the temperature corresponding to $c_V = 2.0$ by linear interpolation of the data. The same method was applied to construct the other lines presented in our work. Shear viscosity was computed by integrating the shear stress autocorrelation function.

We also have calculated maxima of the following quantities: isothermal compressibility $\beta = \frac{1}{\rho}\left(\frac{\partial \rho}{\partial P}\right)_T$, isobaric heat expansion $\alpha = -\frac{1}{\rho}\left(\frac{\partial \rho}{\partial T}\right)_P$ and density fluctuations $\varsigma = \frac{<\Delta N^2>}{N^2} = \left(\frac{\partial \rho}{\partial P}\right)_T$. The maxima of all these quantities rapidly decay with increasing temperature. All three quantities were computed by numerical differentiation of equation of states. The same interpolation method was used to find the maximum of $c_P$. All maxima mentioned above rapidly decay on increasing temperature. However, they can preserve a very small peak which can be formally considered as a maximum. Here we apply the following criteria to find the highest temperature at which a maximum exists: if the ratio between the maximum value and the value of 10 % away is less than 1.01 then we assume that the maximum disappeared. This criterion is useful for both numerical and experimental situation since it corresponds to the limits of numerical or experimental precision in measuring different quantities. Everywhere in text and in the figures, densities, temperatures and pressures of LJ liquid are given in the units $\rho/\rho_c$, $T/T_c$, and $P/P_c$. The following critical parameters, averaged from literature sources, were used for the LJ system: $\rho_c$=0.314, $T_c$=1.31.

The diffusion coefficients have been calculated from the viscosity data using Stokes-Einstein relationship. To prove condition (3) for real liquids we have taken the relationship $V_{sl}$=2.3$V_{th}$ (where the coefficient 2.3 for longitudinal sound velocity instead of the value 2 for shear sound velocity was taken to get matching with the condition (5) for the same substances at $P$~100$P_c$). Other experimental data [$c_v(T,P)$, $c_P(T,P)$, $\rho(T,P)$, etc.] for Ar, Ne, and $N_2$ have been taken directly from NIST database [9].